\documentstyle[preprint,aps,epsfig,amsbsy,floats]{revtex}
 \tightenlines

\begin{document}
\newcommand{\beq}{\begin{equation}}
\newcommand{\eeq}{\end{equation}}
\newcommand{\beqa}{\begin{eqnarray}}
\newcommand{\eeqa}{\end{eqnarray}}
\newcommand{\sr}{\sqrt}
\newcommand{\fr}{\frac}
\newcommand{\mn}{\mu \nu}
\newcommand{\G}{\Gamma}

\draft \preprint{hep-th/0407066,~ INJE-TP-04-05}
\title{Cosmological parameters in noncommutative  inflation}
\author{Yun Soo Myung\footnote{E-mail address:
ysmyung@physics.inje.ac.kr}}
\address{
Relativity Research Center and School of Computer Aided Science\\
Inje University, Gimhae 621-749, Korea} \maketitle

\begin{abstract}
We investigate how the uncertainty of noncommutative spacetime
could explain the WMAP data. For this purpose,  the spectrum is
divided into the IR and UV  region.  We introduce a noncommutative
parameter of $\gamma_0$ in the IR region  and a noncommutative
parameter of $\mu_0$ in the UV region. We calculate cosmological
parameters using the slow-roll expansion in the UV region and a
perturbation method in the IR region. The power-law inflation is
chosen to obtain explicit forms for the power spectrum, spectral
index, and running spectral index. Further, these are used to fit
the data.
\end{abstract}

\thispagestyle{empty}
\setcounter{page}{0}
\newpage
\setcounter{page}{1}

\section{Introduction}
String  theory as a candidate for the theory of everything can say
something  about cosmology\cite{Bran1}. Focusing on a universal
property of string theory, it is very interesting to study its
connection to cosmology. The universal property which we consider
is a stringy spacetime uncertainty relation (SSUR) of $ \triangle
t_p \triangle x_p \ge l^2_s$ where $l_{s}$ is the string length
scale\cite{Yone}. This implies that spacetime is noncommutative.
It is compared to a stringy uncertainty relation of  $ \triangle
x_p \triangle p \ge 1+ l^2_s \triangle p^2$. The former is
considered as a universal property for strings as well as
D-branes, whereas the latter is suitable only for strings.
Spacetime noncommutativity does not affect the evolution of the
homogeneous background. However, this leads to a coupling between
the fluctuations generated in inflation and the flat background of
Friedmann-Robertson-Walker (FRW) spacetime\cite{BH}.

On the other hand, it is generally accepted that curvature
perturbations produced during inflation are considered  to be
 the origin of  CMB anisotopies and inhomogeneities for large-scale structure.
  The
first year results of WMAP put forward more constraints on
cosmological models and confirm the emerging standard model of
cosmology, a flat $\Lambda$-dominated universe seeded by
scale-invariant adiabatic gaussian fluctuations\cite{Wmap1}. In
other words, these results coincide with  predictions of the
slow-roll inflation using a single inflaton.  Also WMAP brings
about some new intriguing results: a running spectral index of
scalar  perturbations and an anomalously low quadrupole of the CMB
power spectrum\cite{Wmap2}. If inflation is affected by physics at
a short distant close to string scale, one expects that the
spacetime uncertainty must be encoded in the CMB power
spectrum\cite{LMMP}. For example, the noncommutative power-law
inflation may produce a negative running spectral index to fit the
data of WMAP\cite{HM1,TMB,HM2}.

In the noncommutative inflation, there exist the two regions: IR
and UV region. The former covers a small energy scale ($k<k_s$),
while the latter covers a large energy scale ($k>k_c$). In the IR
region we expect to find a strongly noncommutative effect on the
inflation but we could not use a conventional approach such as the
slow-roll approximation to obtain cosmological parameters.
Although we find a weakly noncommutative effect in the UV region,
the slow-roll approximation is employed in computing cosmological
parameters. If one chooses the IR critical scale $k_s$ and UV
critical scale $k_c$ appropriately, these parameters could be used
to fit the data. It is necessary to introduce both the IR and UV
regions to cover the whole noncommutative effect on the inflation.

 Recently the UV cosmological parameters have been
calculated  with the slow-roll parameters $\epsilon_1$ and
$\delta_n$ and a noncommutative parameter $\mu_0$\cite{HM3}. It
was shown that  the noncommutative parameter $\mu_0$ could be
regarded as a zeroth order slow-roll parameter in the UV
region\cite{KLM,KLLM}. In this work, we make a further progress in
this direction. We study the IR region  parallel with the UV
region by making use of $\gamma_0$. We show how the uncertainty of
noncommutative spacetime could explain the WMAP data.

 The organization of this work is as follows. In
Section II we review the framework for a perturbative inflation.
In Section III, we calculate cosmological parameters in the UV
region. Section IV is devoted to computing cosmological parameters
in the IR region. Finally we discuss our results in Section V.

\section{framework for inflation}
\subsection{Commutative inflation}
Our starting point is an effective action based on an inflaton
minimally coupled to gravity
\begin{equation}\label{action}
A= \int  d^4 x \sqrt{-g} \left[ -\frac{M^{2}_{P}}{2}R +
\frac{1}{2}(\partial_\mu \phi)^2 - V(\phi) \right],
\end{equation}
where $M^{2}_{P}$ is the reduced Planck mass defined by
$M_{P}=1/(8\pi G)^{1/2}=1/l_P$ and its length scale is given by
$l_P=8.101 \times 10^{-33}{\rm cm}$.  Considering a flat FRW
background of $ds^2_{FRW}=dt^2-a(t)^2 d{\bf x}\cdot d{\bf x}$, one
finds the Friedmann equations \beq \label{Feqn}
H^2=\frac{\rho}{3M_P^2},~~\dot{H}=-\fr{1}{2M_P^2}(\rho+p) \eeq
with the Hubble parameter $H=\dot{a}/a$. From the action
(\ref{action}), one finds the equation
\begin{equation}
\label{seq} \ddot{\phi}+3H\dot{\phi}=-V^{\prime},
\end{equation}
where dot and prime denote   derivative with respect to a cosmic
time $t$ and $\phi$, respectively. Its energy density and pressure
are given by $\rho=\dot{\phi}/2+V$ and $p=\dot{\phi}/2-V$.

We briefly review  the slow-roll approximation. This approximation
means that an inflation $(\ddot a>0)$ is driven by a single scalar
field slowly rolling down its potential toward a local minimum.
Then Eqs.(\ref{Feqn}) and (\ref{seq}) take the following form
approximately:
\begin{equation}
H^2\approx \frac{V}{3M_P^2},~\dot{H}=-\fr{\dot{\phi}^2}{2M_P^2},~
\dot{\phi}\approx -V'/3H.
\end{equation}
 In order to take this approximation into account, we
introduce  slow-roll parameters (H-SR towers) as
\begin{equation}
\epsilon_1 \equiv -\frac{\dot H}{H^2}=\fr{1}{2M_p^2}
\Big(\fr{\dot{\phi}}{H}\Big)^2,~~\delta_n \equiv
\frac{1}{H^n\dot{\phi}}\frac{d^{n+1}\phi}{dt^{n+1}}
\end{equation}
which satisfy the slow-roll condition:
$\epsilon_1<\xi,~|\delta_n|<\xi^n$ for some small perturbation
parameter $\xi$.
 Here
 the subscript denotes  slow-roll (SR)-order in the slow-roll
expansion.
 A scalar  metric perturbation to the flat
FRW background is expressed  in the longitudinal gauge as
\cite{bardeen}
\begin{equation}
\label{con-p}
 ds^2_{con-p} = a^2(\eta) \left\{(1+2A) d\eta^2
-(1+2\psi)d{\bf x}\cdot d{\bf x} \right\}
\end{equation}
with a conformal time $\eta$ defined by $d\eta = dt / a $. We get
a relation of $\psi=A$ because the stress-energy tensor does not
have any off-diagonal component. It is convenient to express the
scalar perturbations in terms of the curvature perturbation ${\cal
R}$~\cite{lukash}
\begin{equation}
{\cal R} = \psi-\frac{H}{ \dot \phi} \delta \phi,
\end{equation}
where $\delta\phi$ is a perturbation of inflaton: $\phi({\bf
x},\eta)=\phi(\eta)+\delta\phi({\bf x},\eta)$.

Introducing a parameter $z$ and a canonical scalar $\varphi$ as
\begin{equation}
z \equiv \frac{a\dot \phi}{H} \ \ \ \mbox{and}\ \ \ \varphi \equiv
a\left(\delta\phi-\frac{\dot\phi}{H} \psi\right) = -z {\cal R},
\end{equation}
the bilinear action for the scalar perturbations leads to a
canonical form \cite{mukhanov}
\begin{equation}
\label{bilinear} S =\frac{1}{2}  \int d\eta\,d^3{\bf x} \left[
\left(\frac{\partial\varphi}{\partial\eta}\right)^2 -
\left(\nabla\varphi\right)^2 +
\left(\frac{1}{z}\frac{d^2z}{d\eta^2}\right)\varphi^2 \right].
\end{equation}
We note that $z$ encodes all information about an expanding
universe of  inflation.  Because the background  is isotropic and
homogeneous, we can expand  $\varphi$  in terms of Fourier modes
as $ \varphi({\bf x},\eta)=\int \fr{d^3{\bf k}}{(2 \pi)^{3/2}}
\varphi_{{\bf k}}(\eta) e^{i{\bf k}\cdot {\bf x}}$ with
$\varphi_{{\bf k}}(\eta)=a_{\bf k}\varphi_k(\eta)
+a^\dagger_{-{\bf k}}\varphi^*_k(\eta)$\cite{LL}. Its
Fourier-transformed action can be expressed in terms of ${\cal
R}_{\bf k}=-\varphi_{\bf k}/z$  as \beq \label{codact} {\cal F}[S]
=\fr{V_T}{2}\int d\eta ~d^3{\bf k}~z^2(\eta)~\Big[\fr{d{\cal
R}_{-{\bf k}}}{d\eta}\fr{d{\cal R}_{\bf k}}{d\eta}- k^2{\cal
R}_{-{\bf k}}{\cal R}_{\bf k}\Big], \eeq where $V_T$ is the total
spatial volume. This action is a useful form because replacing
$z^2$ by $a^2$ in the integrand leads  to that for  the tensor
perturbation and  its form persists to the noncommutative case.
From this action one finds the Mukhanov equation
\begin{equation}
\label{ceqsn} \frac{d^2\varphi_k}{d\eta^2} +
\left(k^2-\frac{1}{z}\frac{d^2z}{d\eta^2}\right)\varphi_k = 0
\end{equation}
which governs the dynamics for  evolution  of a $k$-th scalar mode
($\varphi_k$).

\subsection{Noncommutative inflation}

We introduce another time coordinate $\tau$ defined by $d\tau=adt$
to incorporate the noncommutative spacetime
appropriately\cite{BH}. Its connection to the conformal time is
given by $d\eta/d\tau=1/a^2$. The other noncommutative  approach
appeared in ref.\cite{FKM}. Then Eq.(\ref{con-p}) can be rewritten
as
\begin{equation}
ds^2_{non-p} = a^{-2}(\tau)(1+2A) d\tau^2 - a^2(\tau)(1+2\psi)
d{\bf x}\cdot d{\bf x}.
\end{equation}
Using this time, the SSUR of $ \triangle t_p \triangle x_p \ge
l^2_s$ becomes
\begin{equation}\label{non-c}
\triangle \tau \triangle x \ge l^2_s. \eeq  Considering $\triangle
x \sim 1/k$, a range of the time-uncertainty is given by
$\triangle \tau =l_s^2k$.  We propose the transition to
noncommutative spacetime obeying Eq.(\ref{non-c}) by taking the
operator appearing in the bilinear action in Eq.(\ref{bilinear})
and replacing all multiplications by $*$-products\cite{Cal3}.
Performing the Fourier transform, the SSUR modifies  its action
minimally as  \beq \label{modact}{\cal F}[\tilde{S}] =\fr{V_T}{2}
\int d\tilde{\eta} ~d^3{\bf
k}~z_k^2(\tilde{\eta})~\Big[\fr{d\tilde{{\cal R}}_{-{\bf
k}}}{d\tilde{\eta}}\fr{d\tilde{{\cal R}}_{\bf k}}{d\tilde{\eta}}-
k^2\tilde{{\cal R}}_{-{\bf k}}\tilde{{\cal R}}_{\bf k}\Big]. \eeq
Here $z_k$ a smeared version of $z$ and $\tilde{\eta}$  a new
conformal time are given by \beq \label{zzz}
z^2_k(\tilde{\eta})=z^2y^2_k(\tilde{\eta}),
~~y^2_k=\sqrt{\beta^+_k\beta^-_k}=\fr{a_+^2+a_-^2}{2a_+a_-},
~~\fr{d\tilde{\eta}}{d\tau} =\sqrt{\fr{\beta^-_k}{\beta^+_k}}
\equiv \fr{1}{a^2_{{\rm eff}}}= \fr{1}{a_+a_-}, \eeq where \beq
\label{betak} \beta^{\pm}_k=\fr{1}{2} \left[a^{\pm 2}_++a^{\pm
2}_-\right] \eeq with $a_{\pm}\equiv a(\tau \pm  \triangle \tau)$.
We note that  the SSUR does not affect an evolution of the
homogeneous background. However, this leads to a coupling between
$\tilde{\varphi}_k(\tilde{\eta})$ generated during inflation and
the flat FRW background through $z_k$.  Actually the SSUR  induces
the uncertainty of time in defining $a$. If one takes a limit of
$\triangle \tau \to 0(l_s \to 0)$, one finds $\beta_k^{\pm}\to
a^{\pm 2}$. This implies that the commutative case is recovered
from the noncommutative formalism: $y_k \to 1,~a_{\rm eff}^2 \to
a^2, ~z_k \to z,~ \tilde{\eta}\to \eta,~ d\tilde{\eta}/d\tau \to
d\eta/d\tau,~\tilde{\varphi}_k(\tilde{\eta}) \to \varphi_k(\eta)$.

From Eq.(\ref{modact}), we derive
 the Mukhanov equation for the noncommutative inflation
\begin{equation}
\label{eqsn} \frac{d^2\tilde{\varphi}_k}{d\tilde{\eta}^2} +
\left(k^2-\frac{1}{z_k}\frac{d^2z_k}{d\tilde{\eta}^2}\right)\tilde{\varphi}_k
= 0.
\end{equation}
Our task is to solve Eq.~(\ref{eqsn}) in the UV and IR regions.
The key step to calculate cosmological parameters is to use the
horizon crossing time at $\tilde{\eta}=\tilde{\eta}_*$ in the UV
region and the saturation time at $\tilde{\eta}=\tilde{\eta}_0$ in
the IR region.  For this purpose we consider a power-law inflation
of $a(t)=a_0t^p$ with $p>1$. Taking $a_0=[1/l(p+1)]^p$, we have
$a(\tau)\equiv\alpha_0\tau^{\fr{p}{p+1}}=(\tau/l)^{\fr{p}{p+1}}$
and $H=[p/(p+1)](l^p\tau)^{-\fr{1}{p+1}}$. From $k=a_{\rm
eff}/l_s$, one finds an important relation between $\tau$ and
$k$\cite{HM2} \beq \label{tauk}\tau(k)=\triangle \tau \Big[
1+\Big(\fr{k}{k_s}\Big)^{\fr{2}{p}}\Big]^{\fr{1}{2}} \eeq with a
critical scale $k_s=l_s^{p-1}/l^p=1/(l^pM_s^{p-1})$. This scale is
the ratio of a string scale $l_s$ to other scale  $l$ related to
$a_0$.

\section{UV cosmological parameters}
In this section we compute cosmological parameters in the UV
region.  In order to calculate these, we have to specify both the
time at $\tilde{\eta}=\tilde{\eta}_0$ when the $k$-mode is
generated and the later time at $
\tilde{\eta}=\tilde{\eta}_*>\tilde{\eta}_0$ when it crosses the
Hubble horizon. However, in the UV region, the cosmological energy
scale when a perturbation is generated is much smaller than the
string energy scale $M_s=1/l_s$:
$H(\tilde{\eta}>\tilde{\eta}_0)\le H(\tilde{\eta}_0)\ll M_s$.
Hence the time $\tilde{\eta}_0$ is not crucial because in the UV
region, all modes are generated inside the horizon. Further,
noncommutative effects are soft and thus $a_{\pm}$ could be
Taylor-expanded up to first-order as\cite{Cal3}
 \beq \label{taylor}a(\tau \pm \triangle
\tau)=a(\tau)[1\pm\sqrt{\mu_0} +
\{\pm\sqrt{\mu_0}-(1\pm\sqrt{\mu_0})\ln(1\pm
\sqrt{\mu_0})\}\epsilon_1]+ {\cal O}(\epsilon_1^2) \eeq with a
noncommutative parameter in the UV region\cite{HM3}
\begin{equation}
\mu_0(t,k)=\Big(\fr{\triangle \tau
H}{a}\Big)^2=\Big(\fr{kH}{aM^2_s}\Big)^2.
\end{equation}
 Its total time derivative is given by\cite{KLM,KLLM} \beq \label{muderi}
\fr{d\mu_0}{dt}=-4H\mu_0\epsilon_1.\eeq Hence we interpret $\mu_0$
to be a slow-roll parameters in addition to $\epsilon_1,~\delta_n$
and its subscript denotes the zeroth-order in the slow-roll
expansion. $\mu_0(t,k)$ is a function of $t$ and $k$ at the
beginning, but one finds two interesting forms: $\mu_*(t)
=2(H/M_s)^4$ at pivot scale $k=k_*=\sqrt{2}aH$   and
$\mu_0(k)=(k_c/k)^{4\epsilon_1}$ for power-law inflation. Although
there is no direct relation between $\mu_0$ and the slow-roll
potential $V(\phi)$, we don't doubt that $\mu_0$ is regarded  as a
slow-roll parameter\footnote{ In order to show it, we consider the
horizon-flow approximation\cite{STG} whose parameters are defined
as $\tilde{\epsilon}_0=H_{inf}/H$ and
$\tilde{\epsilon}_{i+1}=d\tilde{\epsilon}_i/dN$ with $H_{inf}$ the
Hubble parameter at some chosen time and  $N=\int Hdt$  the
$e$-folding number since the horizon-crossing time. Here
$\tilde{\epsilon}_0$ is a geometric quantity. One takes this
quantity by hand to obtain higher-order slow-roll parameters
$\tilde{\epsilon}_{i+1}$.   $\mu_0$ arises from an effect of the
noncommutative spacetime and it belongs to a geometric quantity
involving a string scale. Similarly we could include $\mu_0$ as
another slow-roll parameter to carry with a noncommutative
effect.}. At this stage we note a procedure of realizing the
noncommutative effect on the cosmological parameters: the SSUR
$(\triangle \tau =kl_s^2)\to a_{\pm} \to \mu_0(t,k) \to z_k \to$
cosmological parameters.

We start with the slow-roll approximation  to calculate
cosmological parameters. This  means that
$\mu_0,~\epsilon_1,~\delta_1$ are taken to be approximately
constant in calculation of the noncommutative power spectrum.  To
this end we obtain a potential-like term up to first order
 \beq \label{nonzzz}
\frac{1}{z_k}\frac{d^2z_k}{d\tilde{\eta}^2}\simeq
2(aH)^2\Big(1-2\mu_0+\epsilon_1+\fr{3}{2}\delta_1\Big) \eeq and
relations from Eqs.(\ref{zzz}) and (\ref{taylor}) \beq
\label{nonaH}aH\simeq
-\fr{1}{\tilde{\eta}}(1+\mu_0+\epsilon_1),~y_k\simeq 1+\mu_0,~
 \eeq
 Then Eq.(\ref{eqsn}) takes the same form as in the commutative
case\cite{SL}
 \beq
\frac{d^2\tilde{\varphi}_k}{d\tilde{\eta}^2} +
\left(k^2-\frac{(\nu^2-\fr{1}{4})}{\tilde{\eta}^2}\right)\tilde{\varphi}_k
= 0\eeq with the same index $\nu=\fr{3}{2}+2\epsilon_1+\delta_1$
except replacing $\eta$ by $\tilde{\eta}$. Its asymptotic solution
to Eq.(\ref{eqsn}) in the limit of $-k\tilde{\eta} \to \infty$
takes a plane-wave
\begin{equation}\label{nbc}
\tilde{\varphi}_k =\frac{1}{\sqrt{2k}}e^{-ik\tilde{\eta}}.
\end{equation}
In the limit of $-k\tilde{\eta} \to 0$, one finds an asymptotic
form of the Hankel function $H^{(1)}_\nu(-k\tilde{\eta})$ \beq
\label{nonphi}\tilde{\varphi}_k \simeq
e^{i(\nu-\fr{1}{2})}2^{\nu-\fr{3}{2}}\fr{\Gamma(\nu)}{\Gamma(\fr{3}{2})}
\fr{1}{\sqrt{2k}}(-k\tilde{\eta})^{\fr{1}{2}-\nu}. \eeq
 Then the  noncommutative power spectrum is defined  by
\begin{equation}\label{nps}
P^{UV}_{R}(k) = \left(\frac{k^3}{2\pi^2}\right)
\lim_{-k\tilde{\eta}\rightarrow0}\left|\frac{\tilde{\varphi}_k}{z_k}\right|^2.
\end{equation}
One finds a scale-dependent power spectrum
 \beq
P^{UV}_{R}(k)
=\frac{H^4}{(2\pi\dot{\phi})^2}\Big[2^{\nu-\fr{3}{2}}\fr{\Gamma(\nu)}{\Gamma(\fr{3}{2})}\Big]^2
\Big(\fr{k}{aH}\Big)^{-2(2\epsilon_1+\delta_1)}
\fr{1}{(1+\mu_0+\epsilon_1)^{2(1+2\epsilon_1+\delta_1)}(1+\mu_0)^{2}}.
\eeq which leads to by making use of  the Taylor expansions
 \beq \label{nonps1} P^{UV,1st}_{R}(k)
= \frac{H^4}{(2\pi\dot{\phi})^2} \left\{ 1 -4\mu_0-2\epsilon_1  +2
\left(\alpha-\ln\Big(\fr{k}{aH}\Big)\right)(2\epsilon_1+\delta_1)
\right\}.  \eeq  up to first order in slow-roll parameters. Here
$\alpha=2-\ln2-\tilde{\gamma}=0.729637$.  In the limit of $\mu_0
\to 0$, $P^{UV,1st}_{R}(k)$ reduces to the commutative power
spectrum\cite{MS1,STW}, while in the extreme slow-roll limit of
$\epsilon_1,~\delta_1 \to 0$, one finds the de Sitter result
including $\mu_0$\cite{HM3}. In the noncommutative  approach the
horizon crossing occurs at
$k^2=\frac{1}{z_k}\frac{d^2z_k}{d\tilde{\eta}^2}$\cite{BH}. Hence,
from  Eq.(\ref{nonzzz}) we use the  pivot scale $k_{*}=
\sqrt{2}aH$ instead of a pivot scale of $k_*=aH$ for a commutative
inflation. Finally, we obtain the noncommutative power spectrum up
to first order as
 \beq \label{nonps11}P^{UV,1st}_{R}(k)
=\left. \frac{H^4}{(2\pi\dot{\phi})^2} \left\{ 1 -2\epsilon_1
-4\mu_0+ 2\alpha_*(2\epsilon_1+\delta_1) \right\}\right|_{k=k_*}
\eeq with $\alpha_*=\alpha-\ln2/2$. Let us compare
Eq.(\ref{nonps11}) with the commutative power spectrum. A change
of the pivot scale from $k_*=aH$ to $k_{*}=\sqrt{2}aH$ amounts to
replacing $\alpha=0.7296$ by $\alpha_*=0.3831$ in the first-order
calculation\cite{STW}.  Thus  the SSUR  imprints  on cosmological
parameters by means  of $\alpha \to \alpha_*$ and
$\mu_0\not=0$\cite{KLM,KLLM}.

In order to calculate the power spectrum even for  first-order
correctly, one has to use the slow-roll expansion based on Green's
function technique\cite{SG,KLM1,KM}. The key step in the slow-roll
expansion is to use Eq.(\ref{muderi}) in deriving the power
spectrum. In the case of $\mu_0$=0, the slow-roll approximation
and slow-roll expansion give the same power spectrum up to first
order. However, in the case  of $\mu_0 \not=0$, two provide
different results. The details appeared in ref.\cite{KLM,KLLM}.
The slow-roll approximation is not generally suitable for the
noncommutative case. From now on we obtain cosmological parameters
using the slow-roll expansion.  Also we wish to compare the
noncommutative cosmological parameters with the WMAP data.  As an
example, we choose the power-law inflation like
$a(t)=a_0t^p,~H=p/t,~z=a\sqrt{2/p}M_P$ whose potential is given by
\beq V(\phi)=V_0
\exp\Big(-\sqrt{\fr{2}{p}}\fr{\phi}{M_P}\Big).\eeq Then slow-roll
parameters are given  by\footnote{From Eq.(\ref{tauk}), we have
$\tau \simeq l(kl_s)^{\fr{p+1}{p}}$ in the UV region of $k>k_s$.
Then we obtain a relation $k \simeq
\fr{1}{l_s}\Big[\fr{p}{lH(p+1)}\Big]^p$. With
$k_c=\Big[\fr{p(2p-1)}{(p+1)^2}\Big]^{\fr{p+1}{4}} k_s$, one finds
$(k_c/k)^{\fr{4}{p}} \simeq 2(H/M_s)^4$ for a large $p$ which
satisfies $p \pm 1 \simeq p$.} \beq \label{pislp} \mu_0(k)\equiv
\Big(\fr{k_c}{k}\Big)^{\fr{4}{p}}\simeq 2\Big(\fr{H}{M_s}\Big)^4
\equiv \mu_*(t), ~\epsilon_1=\fr{1}{p},~\delta_1=-\fr{1}{p},~
\delta_2=\fr{2}{p^2},~\delta_3=-\fr{6}{p^3}
,~\delta_4=\fr{24}{p^4}\eeq where $\mu_0(k)$ is given by a
solution to $d\mu_0/d \ln k=-(4/p)\mu_0$. Also $\mu_*(t)$
satisfies Eq.(\ref{muderi}). A UV critical scale $k_c$ is given by
$k_c\simeq 2^{p/4}k_s$ approximately. A UV region of $\mu_0< 1(k_c
< k)$ means $\mu_*<1(H < M_s)$. In the UV region, we calculate
power spectrum using $H(\tilde{\eta}_*)$ which is a solution to
$\mu_0(k=k_*)=\mu_*(\tilde{\eta}=\tilde{\eta}_*)$. Then the
noncommutative power spectrum takes the form in the slow-roll
expansion\beq \label{pips1}\tilde{P}^{UV,1st}_{R}(k) =\left.
P^{c,1st}_{R}(k)+\frac{\mu_0(k)H^2}{(2\pi z)^2} \left\{
       -4  +\fr{12(1-2\alpha_*)}{p}\right\}\right|_{k=k_*},
 \eeq where the commutative spectrum is given by
 \beq
 \label{2ndpss}
\tilde{P}^{c,1st}_{R}(k) = \left.\frac{H^2}{(2\pi z)^2} \left\{ 1
+\fr{2(\alpha_*-1)}{p}\right\}\right|_{k=k_*}. \nonumber \eeq
Comparing   with Eq.(\ref{nonps11}), the last term  in
Eq.(\ref{pips1}) is new.
 The noncommutative spectral index can be
easily calculated up to second-order
\begin{equation} \label{pisi1}
\tilde{n}^{UV}_{s}(k) = \left.n^{c}_{s}(k)+\mu_0(k)
\left\{\fr{16}{p} + \fr{64\alpha_*}{p^2}\right\}\right|_{k=k_*}
\end{equation}
with the commutative contribution \beq \label{pisi11}
n^{c}_{s}(k)=1-\fr{2}{p}-\fr{2}{p^2}. \eeq Here one finds the last
term in Eq.(\ref{pisi1}) from the slow-roll expansion. Finally the
running spectral index is found to be \beq
\label{pirsi1}\frac{d\tilde{n}^{UV}_s}{d\ln k} = \left.\frac{d
n^{c}_s}{d\ln k} -\mu_0(k)
\left\{\fr{64}{p^2}+\fr{8(32\alpha_*+8)}{p^3}\right\}\right|_{k=k_*},\eeq
but the commutative contribution is zero up to third-order, \beq
\label{pirsi11}\frac{d n^{c}_s}{d\ln k}=0. \eeq The last term in
Eq.(\ref{pirsi1}) comes from the slow-roll expansion.

 We obtain a UV critical scale $k_c=0.998
\times 10^{-5}{\rm Mpc}^{-1}$ and a IR critical scale $k_s=1.05
\times 10^{-6}{\rm Mpc}^{-1}$ by choosing  $l_s=3.49 \times
10^{-29}{\rm cm}$, $l=1.19 \times 10^{-24}{\rm cm}$, and
$p=13$~\footnote{ In this work, we consider only an integer  $p$.
The two scales depend on $p$ critically. For $p=12$, we have large
energy scales of $k_c=0.28{\rm Mpc}^{-1},~k_s=0.035{\rm
Mpc}^{-1}$, whereas for $p=14$, we have small energy scales of
$k_c=3.48\times 10^{-10}{\rm Mpc}^{-1},~k_s=0.28=3.07\times
10^{-11}{\rm Mpc}^{-1}$. All of these are not suitable for
describing a noncommutative inflation.}\cite{HM2,HM3}. For
simplicity we choose $k_c=10^{-5}{\rm Mpc}^{-1}$ and
$k_s=10^{-6}{\rm Mpc}^{-1}$. In this case the critical scale $k_t$
is chosen by $k_t=10^{-3}{\rm Mpc}^{-1}$ which is slightly larger
than $H_0=4.6\times 10^{-4}{\rm Mpc}^{-1}$.  The relevant pivot
scales in the UV region should satisfy $k_* \gg k_c $ and
$k_*>k_t$. Hence we choose $k_*=0.05{\rm Mpc}^{-1}$ for a small
length scale and $k_*=0.002{\rm Mpc}^{-1}$ for a large length
scale to compare with the WMAP data. In this case $\mu_0=0.07275$
at $k_*=0.05{\rm Mpc}^{-1}$ and $\mu_0=0.19588$ at $k_*=0.002{\rm
Mpc}^{-1}$. We trust more the data at $k_*=0.05{\rm Mpc}^{-1}$
than that at $k_*=0.002{\rm Mpc}^{-1}$ because of $\mu_0$ as a
slow-roll parameter should be comparable with $\epsilon_1<0.08$ in
the UV region\cite{HM3}. Here our power spectrum normalization
${\cal A}$ at $k_*=0.05{\rm Mpc}^{-1}$ is defined by
$P_{R}^{UV}=\Big(\frac{aH}{2\pi z}\Big)^2\times {\cal A}
=1.69\times10^{-9}\times {\cal A}$ with ${\cal A}=0.629$, while
the WMAP provides $P_{R} =2.95\times10^{-9}\times {\cal A}$ with
${\cal A}=0.833^{+0.086}_{-0.083}$\cite{Wmap1}. Approximately,
there exists a difference of ``2" in power spectrum normalization
in $10^{-9}$ order. For reference, we have $P_{R}^{UV}=4.55\times
10^{-10}$ at $k_*=0.002 {\rm Mpc}^{-1}$. In Table I, we show
noncommutative spectral index at two different scales up to
first-order and second-order. Noncommutative running spectral
index at two different scales up to second-order and third-order
appear in Table II. A negatively large running spectral index
could be obtained even in the UV  approach.

 \begin{table}
 \caption{UV spectral index and WMAP data at two different scales.}
 \begin{tabular}{c|c|c|c}
 scale (${\rm Mpc}^{-1}$)   & first-order ($n_s^{1st}$) & second-order ($n_s^{2nd}$) & WMAP \\ \hline
 $k=0.05$     & 0.935692   & 0.934413
 &$0.93^{+0.03}_{-0.03}$\\
 $k=0.002$     & 1.08724   & 1.10382
 &$1.20^{+0.12}_{-0.11}$
 \end{tabular}
 \end{table}
 \begin{table}
 \caption{UV running spectral index and WMAP data at two different scales.}
 \begin{tabular}{c|c|c|c}
 scale (${\rm Mpc}^{-1}$)   & second-order ($dn_s^{1st}/d \ln k$) & third-order ($dn_s^{2nd}/d \ln k$) & WMAP \\ \hline
 $k=0.05$     & $-0.0275503$   & $-0.0329171$
 &$-0.031^{+0.016}_{-0.017}$\\
 $k=0.002$     & $-0.0741794$   &$-0.0886296$
 &$-0.077^{+0.05}_{-0.052}$
 \end{tabular}
 \end{table}

\section{IR cosmological parameters}
In the IR region, the situation is quite different from the UV
case\cite{BH,TMB,HM1,Cai}. The perturbed modes are generated
outside the Hubble horizon. Their magnitude depends on the time
when they are generated because they are frozen as soon as  they
are generated. When the SSUR is saturated, this corresponds to the
time $\tilde{\eta}=\tilde{\eta}_0$ ($\tau=\tau_0$).  Actually a
saturation time $\tilde{\eta}_0$ in the IR region plays a similar
role of a pivot scale $\tilde{\eta}_*$ in the UV region. Further,
the perturbed modes start out with their vacuum amplitude.
  In the
case of $\tau \simeq \triangle \tau$ for $k<k_s$, we have \beq
H=\fr{p}{p+1}\Big(\fr{1}{l}\Big)^{\fr{p}{p+1}}\Big(\fr{1}{\tau}\Big)^{\fr{1}{p+1}}
\simeq \fr{p}{p+1}
\Big(\fr{1}{l}\Big)^{\fr{p}{p+1}}\Big(\fr{1}{l_s^2k}\Big)^{\fr{1}{p+1}}.
\eeq One finds for $k<k_s,~ p+1\simeq p$
 \beq
 \gamma_0(k) \equiv \Big(\fr{k}{k_s}\Big)^{\fr{2}{p}}=
 \Big(\fr{p}{p+1}\Big)^{\fr{2(p+1)}{p}}
\Big(\fr{M_s}{H}\Big)^{\fr{2(p+1)}{p}} \simeq
\Big(\fr{M_s}{H}\Big)^2 \equiv\tilde{\gamma}_0(t) \eeq where
 a newly noncommutative
zeroth-order parameter $\gamma_0(k)(\tilde{\gamma}_0(t))$ is
suitable for describing the IR region of either
$\gamma_0(k)<1(k<k_s)$ or
$\tilde{\gamma}_0=\sqrt{2/\mu_*}<1(H>M_s)$. We use a relation of
$\gamma_0(k=k_0)=\tilde{\gamma}_0\tau=\tau_0)$ to obtain
$H(\tau_0)$. Then Eq.(\ref{tauk}) leads to
 \beq \tau=\triangle \tau \Big[
1+\gamma_0\Big]^{\fr{1}{2}}\eeq where at the IR end of $\gamma_0
\to 0$, one has $\tau_0 \to \triangle\tau$, as was shown
previously. It seems that one is not easy to achieve this limit
with $k_s=10^{-6}{\rm Mpc}^{-1}$ because $\gamma_0=0.70$ at
$k=10^{-1}k_s$, $\gamma_0=0.34$ at $k=10^{-3}k_s$, $\gamma_0=0.17$
at $k=10^{-5}k_s$, and $\gamma_0=0.08$ at $k=10^{-7}k_s$. However,
the last corresponds to a length scale of $l=10^{-29}{\rm cm}$
slightly smaller than the string scale $l_s=3.49\times
10^{-29}{\rm cm}$. We define the IR region  to be $\gamma_0\le
0.12$ which is equivalent to $k\le 10^{-6}k_s$ for clarity. In the
IR region we have $k_0=\tau_0
M_s^2=aH(M_s/H)^2=aH\tilde{\gamma}_0=k_*\tilde{\gamma}_0/\sqrt{2}<k_*$,
which is smaller than the corresponding UV pivot scale
$k_*=\sqrt{2}aH$. In the IR region, we have $
a=\alpha_0\tau^{1-\fr{1}{p+1}}\simeq H\tau =H\triangle \tau
\sqrt{1+\gamma_0}$ with $H\simeq \alpha_0 \tau^{-\fr{1}{p+1}}$.
Furthermore, we have $a_{\pm}\simeq H\triangle
\tau(\sqrt{1+\gamma_0}\pm 1)$. From
$\tilde{\gamma}_0(t)=(M_s/H)^2$ for $H<M_s$, we find a relation
\beq \dot{\tilde{\gamma}_0}=2H\tilde{\gamma}_0 \epsilon_1. \eeq
Similarly from $\gamma_0(k)=(k/k_s)^{2/p}$ for $k<k_s$, one finds
$d\gamma_0/d\ln k =(2/p)\gamma_0$. These show that $\gamma_0$
could be treated as a zeroth-order parameter to describe the IR
region. Hereafter in order to evaluate all of cosmological
parameters, we have to use a pivot scale $k=k_0$ for IR  instead
of a pivot scale $k=k_*$ for UV in the previous section. Then the
noncommutative power spectrum from Eqs.(\ref{nps}) and
(\ref{nonphi}) takes the form\cite{Cal3}\beq
\label{npips1}P^{IR}_{R}(k) =\left.
\fr{k^3}{2\pi^2}\fr{1}{2k}\Big(\fr{-1}{k\tilde{\eta}}\Big)^2
\fr{1}{z^2y_k^2}=  \Big(\fr{aH}{2\pi z}\Big)^2
\fr{\gamma_0^3}{(1+\gamma_0)^2(2+\gamma_0)} \simeq \frac{
H^4}{(2\pi)^2\dot\phi^2}
\Big[\fr{\gamma_0^3(k)}{2}\Big]\right|_{k=k_0} \eeq
 with
$\tilde{\eta}=-(1/aH)(a/a_{\rm eff})^2$. Here we find that the IR
power spectrum is the  product of the commutative contribution by
$\gamma_0^3/3$. The latter can be interpreted as a normalization.
For $\gamma_0= 0.12(k_0=10^{-6}k_s)$ and $p=13$, we have $(aH/2\pi
z)^2=7.39\times 10^{-8}$ and $\gamma_0^3/2=8.6 \times10^{-4}$.
Then we have $P^{IR}_{R}=6.3 \times 10^{-11}$ at $k_0=10^{-12}{\rm
Mpc}^{-1}$ which is smaller than $P^{UV}_{R}=1.198 \times 10^{-9}$
at $k_*=0.05{\rm Mpc}^{-1}$. The noncommutative spectral index can
be easily calculated as
\begin{equation} \label{npisi1}
n^{IR}_{s}(k) =\left.
n^{c}_{s}(k)+\Big[\fr{4(3+2\gamma_0)}{(2+\gamma_0)(1+\gamma_0)}\Big]
\epsilon_1\right|_{k=k_0}
\end{equation}
with the commutative contribution \beq \label{npisi11}
n^{c}_{s}(k)=1-4\epsilon_1-2\delta_1=1-\fr{2}{p}. \eeq Also the
last term in Eq.(\ref{npisi1}) could be approximated as
$(6-5\gamma_0)\epsilon_1$. From the above one finds \beq
n^{IR}_{s}(k) \simeq \left.1+\fr{4}{p}\Big(1-\fr{5}{4}\gamma_0(k)
\Big)\right|_{k=k_0} ,\eeq where we find a blue spectral index of
$n_s>1$ for $\gamma_0<4/5$ in the IR region. For $\gamma_0= 0.12$
and $p=13$, this leads to a blue spectral index of
$n_s^{IR}=1.26$.

The IR running spectral index is found to be \beq
\label{npirsi1}\frac{dn^{IR}_s}{d\ln k} = \left. \frac{d
n^{c}_s}{d\ln
k}+(6-5\gamma_0)\epsilon_1[(2-\bar{\sigma})\epsilon_1+2\delta_1]\right|_{k=k_0}
\eeq with \beq
\bar{\sigma}=\fr{2\gamma_0(5+6\gamma_0+2\gamma_0^2)}{(1+\gamma_0)(2+\gamma_0)(3+3\gamma_0)}
\simeq \fr{5}{3}\gamma_0. \eeq Here the commutative contribution
is zero up to third order, \beq \label{npirsi11}\frac{d
n^{c}_s}{d\ln k}=0. \eeq Finally we have \beq \frac{dn^{IR}_s}{d
\ln k} \simeq \left. \Big(-10+ \fr{25}{3}\gamma_0(k)
\Big)\fr{\gamma_0(k)}{p^2}\right|_{k=k_0} .\eeq For $\gamma_0<1$
IR region, we find a negative running spectral index. For
$\gamma_0= 0.12$ and $p=13$, this leads to a negatively small
running spectral index of $\frac{dn^{IR}_s}{d \ln k}=-0.0064$.

\section{Discussion}
\begin{table}
 \caption{ Comparison between UV and IR for power-law inflation $a(t)\equiv a_0t^p=[t/l(p+1)]^p$
           for a large $p$ such  that  $p\pm1\simeq p$.
            Here we distinguish between  a length scale $l$ and string scale $l_s=1/M_s$.}
 \begin{tabular}{c|c|c|c|c}
 region   &SR parameter & critical scale & pivot scale & condition\\ \hline
 UV    & $\mu_0=(k_c/k)^{4/p}\simeq 2(H/M_s)^4=\mu_*$  & $k_c=2^{p/4}k_s$ &$k_*=\sqrt{2}aH$&$k_c<k,~H<M_s$\\
 IR   & $\gamma_0=(k/k_s)^{2/p}\simeq (M_s/H)^2=\tilde{\gamma}_0$ & $k_s=l^{p-1}_s/l^p$ &$k_0=aH\gamma_0$&$k_s>k,~H>M_s$ \\
 \end{tabular}
 \end{table}

We summarize important information about UV and IR cases in Table
III. In the UV region with the power-law inflation,  we mainly use
a noncommutative parameter $\mu_0(k)$ to compute cosmological
parameters at the pivot scale $k=k_*=\sqrt{2}aH > k_c$ with a UV
critical scale $k_c$. $\mu_*(t)$ is employed only to calculate the
UV power spectrum of $(aH/2\pi z)^2$ at the horizon-crossing time
$\tilde{\eta}=\tilde{\eta}_*$ with $H$. On the other hand, in the
IR region, we mainly use a noncommutative parameter $\gamma_0(k)$
to compute cosmological parameters at the pivot scale
$k=k_0=\tau_0M_s^2=aM_s^2/H < k_s$ with a IR critical scale $k_s$.
$\tilde{\gamma}_0(t)$ is employed only to calculate the IR power
spectrum of $(aH/2\pi z)^2 $ at the saturation time
$\tilde{\eta}=\tilde{\eta}_0$.

At the pivot scale $k_*=0.05{\rm Mpc}^{-1}$, the UV power spectrum
is given by $P_R^{UV}= 1.198 \times 10^{-9}$ and $P_R^{WMAP}\simeq
2.46 \times 10^{-9}$ from the WMAP data, while  $P_R^{UV}= 4.55
\times 10^{-10}$ at $k_*=0.002{\rm Mpc}^{-1}$. At the very small
pivot scale $k_0=10^{-12}{\rm Mpc}^{-1}$, the IR power spectrum is
$P_R^{IR}= 6.3 \times 10^{-11}$. In general the power spectrum
decreases when the energy scale $k$ decreases. The discrepancy
with the data is not so important because the normalization of the
noncommutative power spectrum depends on the string length scale
$l_s$ and a scale $l$ related to $a_0$. Concerning  the
second-order spectral index, we have a red one of
$n_s^{UV}=0.935692<1$ at $k_*=0.05{\rm Mpc}^{-1}$ and a blue one
$n_s^{UV}=1.10382>1$ at $k_*=0.002 {\rm Mpc}^{-1}$. Fortunately
these are close to the data of $0.93^{+0.03}_{-0.03}$ at
$k_*=0.05{\rm Mpc}^{-1}$ and $1.20^{+0.12}_{-0.11}$ at
$k_*=0.002{\rm Mpc}^{-1}$.  At the very small pivot scale
$k_0=10^{-12}{\rm Mpc}^{-1}$, the IR spectral index is  largely
blue ($n_s^{IR}= 1.26$). Here we find that the spectral index
increases when the energy scale $k$ decreases. For the third-order
running spectral index, we have a negative one of $dn_s^{UV}/d\ln
k=-0.0329171$ at $k_*=0.05{\rm Mpc}^{-1}$ and a negative one
$dn_s^{UV}/d\ln k=-0.086296$ at $k_*=0.002 {\rm Mpc}^{-1}$. These
are close to the data of $-0.031^{+0.016}_{-0.017}$ at
$k_*=0.05{\rm Mpc}^{-1}$ and $-0.077^{+0.05}_{-0.052}$ at
$k_*=0.002{\rm Mpc}^{-1}$. But the IR spectral index is a
negatively small  quantity ($dn_s^{IR}/d\ln k= -0.0064$) at the
very small pivot scale $k_0=10^{-12}{\rm Mpc}^{-1}$.

In the IR end ($k \ll k_s$) with a power-law inflation, one finds
the scale-dependent power spectrum $P(k) \approx k^{4/(p+1)}$
which implies a blue spectral index
$n_s=1+4/(p+1)>1$\cite{TMB,Cai}. In the UV end ($k \gg k_c$) one
finds $P(k) \approx k^{-2/(p-1)}$ which implies a red spectral
index $n_s=1-2/(p-1)<1$. In this work we find $n_s^{UV} \to 1-2/p
$ in the limit of $\mu_0 \to 0$ and $n_s^{IR} \to 1+4/p $ in the
limit of $\gamma_0 \to 0$. Considering a large $p$ such as $p\pm 1
\simeq p$, the two results are nearly the same.

In conclusion, we show that the unfamiliar IR region is treated as
in the familiar UV region by introducing a noncommutative
parameter $\gamma_0(\tilde{\gamma}_0)$. But at the scales of $l_s
\sim 10^{-29}{\rm cm}$ and $l \sim 10^{-24}{\rm cm}$, the IR
region of $k<k_s=1.05\times 10^{-6}{\rm Mpc}^{-1}$  is too small
to cover the cosmologically relevant scales of $10^{-4}{\rm
Mpc}^{-1}< k <10^{-1}{\rm Mpc}$. In this case the UV region is
relevant to fitting the data, instead of the IR region. The other
choice of the string scale  was introduced to show the relevance
of the IR region\cite{TMB}.

\subsection*{Acknowledgements}
I thank Hungsoo Kim and Hyung Won Lee for helpful discussions.
This was supported in part by KOSEF, Project No.
R02-2002-000-00028-0 and Grant from Inje University.

\end{document}